\documentclass[a4paper,11pt]{article}
\pdfoutput=1 % if your are submitting a pdflatex (i.e. if you have
\usepackage{jheppub} % for details on the use of the package, please
\usepackage[utf8]{inputenc} % if needed
\usepackage[T1]{fontenc} % if needed
\usepackage{feynmp}
\usepackage{xspace}
\title{Master integrals for two-loop $C$-odd contribution to $e^+e^-\to \ell^+\ell^-$ process.}
\author{Roman N. Lee and Kirill T. Mingulov}
\affiliation{The Budker Institute of Nuclear Physics,\\ 630090, Novosibirsk}
\emailAdd{r.n.lee@inp.nsk.su}

\abstract{We calculate two-loop master integrals for the process of heavy lepton pair production in $e^+e^-$ collisions. We consider the $C$-odd diagrams with three photons in the intermediate state and evaluate the corresponding families of the master integrals in the limit of zero electron mass. Our results for the master integrals are directly applicable in the whole physical region corresponding to the annihilation channel.}

\newcommand{\LiteRed}{\texttt{LiteRed}\xspace}
\newcommand{\Pexp}{\mathrm{Pexp}}
\newcommand{\II}{\mathrm{II}}

\allowdisplaybreaks
\begin{document}
\maketitle

\section{Introduction}

NNLO corrections to the differential cross sections of various QED and QCD processes are now one of the hot topics. One reason for this is the growing precision of the collider experiments which require the corresponding accuracy of the theoretical predictions. This is especially important for the continuing searches of the deviations from Standard Model. 
On the other hand, the present multiloop techniques seem to be able to stand the challenge of evaluation of complicated integrals depending on several kinematic variables, at least, to some extent. The multiloop corrections to the processes involving massive particles are known to be especially complicated. Already at two loops the internal massive lines can prohibit expressing the result in terms of the generalized polylogarithms. However, even in the cases when the results can be written via polylogarithmic functions and constants, the complexity of the massive integral families tends to be higher than that of the massless families. One reason for this is that massive threshold singularities usually involve square roots. The more interesting it is to investigate two-loop multiscale integral families cases by case. From the point of view of developing the multiloop integration technique, these investigations contribute to database of known results and gradually improve our understanding of the functions involved in multiscale diagrams.

Probably, the most appropriate technique for the evaluation of the multiloop multiscale integrals is the method of differential equations \cite{Kotikov1991b,Kotikov1991,Kotikov1991a,Remiddi1997}. Within this approach, one applies the IBP reduction \cite{Tkachov1981,ChetTka1981} to obtain the system of linear ordinary differential equations for the master integrals. Equipped with proper boundary conditions, these equations totally determine the master integrals. A few years ago a remarkable observation has been made in Ref. \cite{Henn2013}. It appeared that in many cases the differential systems for multiloop integrals can be transformed to a form which is perfectly fitted for the calculation of the $\epsilon$ expansion ($\epsilon=2-d/2$ is the parameter of dimensional regularization).

In the present paper we apply the differential equations method to the calculation of the master integrals relevant for the differential cross section of $e^+e^-\to \ell^+\ell^-$ process.
%TODO: check and discuss.
One physical motivation for this calculation is the expected dramatic increase of statistics for taus at Belle II \cite{Kou:2018nap} and Super Charm-Tau factory \cite{Bondar:2013cja}.
In our calculation we neglect the electron mass, while keeping the full dependence on the mass of final particles, and consider the topologies with three photons in the intermediate state. 
These diagrams interfere with the Born diagram in the total cross section. We use the reduction algorithm introduced in Ref. \cite{Lee2014} and further extended in Refs. \cite{Lee2017a,Blondel:2018mad} to reduce the differential systems in $s/m^2$ and $t/m^2$ to $\epsilon$-form and express the solution in terms of the Goncharov's polylogarithms.

Recently, in Refs. \cite{Mastrolia:2017pfy,DiVita2018} similar families of the integrals were calculated (in fact, including the families that we do not consider in the present paper). The results of these papers are also expressed in terms of the Goncharov's polylogarithms. However, these results were obtained at $t<0\&s<0$ and in order to obtain the master integrals in physical kinematic region, corresponding to annihilation channel, one has to perform the analytic continuation. This task appears to be very nontrivial as we will illustrate in the last section.

Our main goal in the present paper was to obtain results which are applicable in the annihilation channel. Our approach differs in several points from the one in  Refs. \cite{Mastrolia:2017pfy,DiVita2018}, namely, in the method of reducing the differential system, in choosing the variables and the point for fixing the boundary conditions. Thanks to the latter, we obtain the expressions for the master integrals which can be immediately used in the whole kinematic region, corresponding to the annihilation channel. They also have functional form, different from Refs. \cite{Mastrolia:2017pfy,DiVita2018}, in terms of the arguments and letters of the polylogarithmic functions and therefore will constitute a nontrivial crosscheck of the results of Refs. \cite{Mastrolia:2017pfy,DiVita2018} once the question of analytical continuation is settled.

\section{Details of the calculation}
\begin{figure}
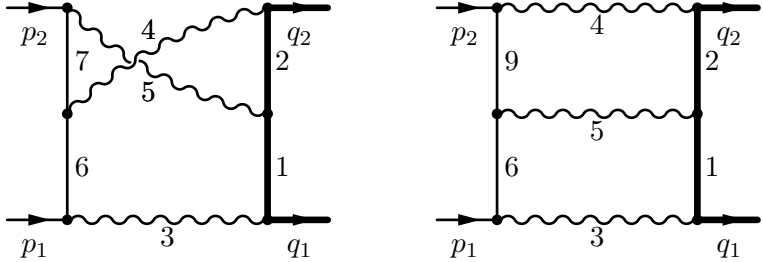
% top topologies
	\centering
%	\begin{fmffile}{FDtopos}
%		\fmfset{arrow_len}{3mm}
%		\fmfset{thick}{2.5thin}
%		\fmfset{dot_size}{1.2thick}
		\parbox{5cm}{
			\centering
			\makebox[0pt][l]{\input{FDtopos.t1}}\includegraphics{FDtopos.1}
%			\begin{fmfgraph*}(150,80)
%				\fmfleft{l1,l2}
%				\fmfright{r1,r2}
%				\fmf{fermion,la=$p_1$}{l1,v1}
%				\fmf{fermion,la=$p_2$}{l2,v2}
%				\fmf{photon,tension=0.3,l.d=1mm,la=$3$}{v1,v3}
%				\fmf{phantom,tension=0.3}{v2,v4}
%				\fmf{fermion,w=thick,la=$q_1$}{v3,r1}
%				\fmf{fermion,w=thick,la=$q_2$}{v4,r2}
%				\fmffreeze
%				\fmf{plain,l.d=1mm,la=$6$}{v1,v5}
%				\fmf{plain,l.d=1mm,la=$7$}{v5,v2}
%				\fmf{plain,w=thick,l.d=1mm,la=$1$}{v3,v6}
%				\fmf{plain,w=thick,l.d=1mm,la=$2$}{v6,v4}
%				\fmffreeze
%				\fmf{photon,left=0.15,rubout=thick,l.d=1mm,la=$4$}{v5,v4}
%				\fmf{photon,right=0.15,rubout=thick,l.d=1mm,la=$5$}{v2,v6}
%				\fmfdotn{v}{6}
%			\end{fmfgraph*}
			\vspace{1mm}\\
%			\centering $j_1=\mathrm{Sr}_1$
		} 
		\hspace{4mm}
		\parbox{5cm}{
			\centering
			\makebox[0pt][l]{\input{FDtopos.t2}}\includegraphics{FDtopos.2}
%			\begin{fmfgraph*}(150,80)
%				\fmfleft{l1,l2}
%				\fmfright{r1,r2}
%				\fmf{fermion,la=$p_1$}{l1,v1}
%				\fmf{fermion,la=$p_2$}{l2,v2}
%				\fmf{photon,tension=0.3,l.d=1mm,la=$3$}{v1,v3}
%				\fmf{photon,tension=0.3,l.d=1mm,la=$4$}{v2,v4}
%				\fmf{fermion,w=thick,la=$q_1$}{v3,r1}
%				\fmf{fermion,w=thick,la=$q_2$}{v4,r2}
%				\fmffreeze
%				\fmf{plain,l.d=1mm,la=$6$}{v1,v5}
%				\fmf{plain,l.d=1mm,la=$9$}{v5,v2}
%				\fmf{plain,w=thick,l.d=1mm,la=$1$}{v3,v6}
%				\fmf{plain,w=thick,l.d=1mm,la=$2$}{v6,v4}
%				\fmffreeze
%				\fmf{photon,l.d=1mm,la=$5$}{v5,v6}
%				\fmfdotn{v}{6}
%			\end{fmfgraph*}
			\vspace{1mm}\\
%			\centering $j_2=\mathrm{Sr}_2$
		}
%	\end{fmffile}
	\caption{The two distinct diagrams contributing to $C$-odd 1p-irreducible NNLO correction to $e^+e^-\to\ell^+\ell^-$. Numbers next to lines correspond to the numeration of the denominators as given in Eq. \eqref{eq:basis}.\label{fig:topologies}}
\end{figure}

\begin{figure}
	\includegraphics[width=1\linewidth]{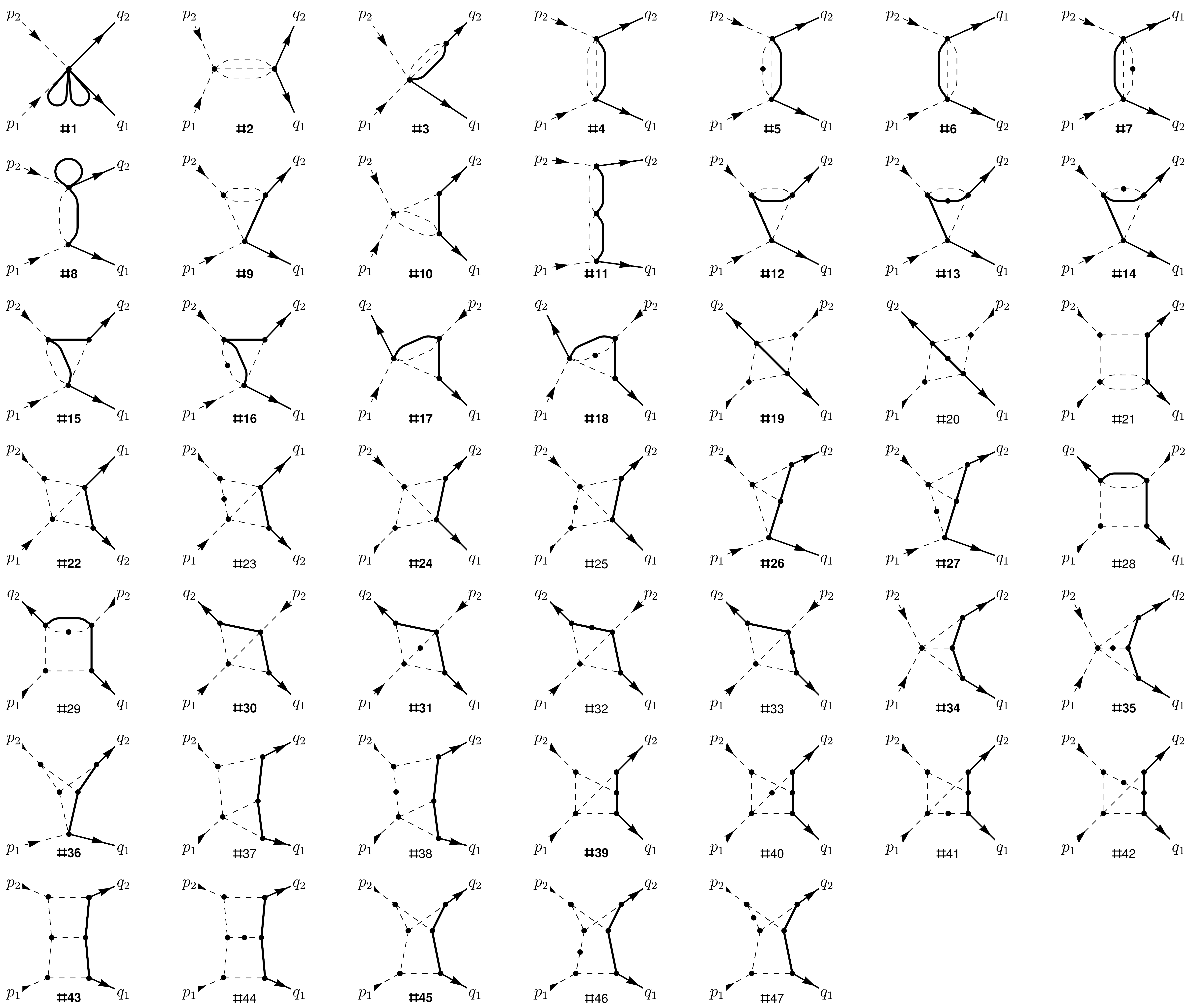}
	\caption{The 47 master integrals for the $\mathtt{tb}$ basis. The 31 master integrals for backward direction are chosen as \#\#1--19,22,24,26,27,30,31,34--36,39,43,45. Their numbers are given in bold on the picture.\label{fig:masters}}
\end{figure}

We consider two topologies depicted in Fig. \ref{fig:topologies}.
Both these topologies can be embedded in one \LiteRed basis defined as 
\begin{gather}\label{eq:basis}
	j(\mathtt{tb},n_1,\ldots n_9) = e^{2\epsilon \gamma_E}\int \frac{d^dl_1}{i\pi^{d/2}}\frac{d^dl_2}{i\pi^{d/2}}\prod_{k=1}^{9}(D_k-i0)^{-n_k}\,\\
	D_1=m^2-\left(l_1+l_2-q_2\right)^2,\ D_2=m^2-\left(l_2-q_2\right)^2,\ D_3=-\left(p_1+p_2-l_1-l_2\right)^2,\ D_4=-l_2{}^2,\nonumber\\
	D_5=-l_1^2,\, D_6=-\left(p_2-l_1-l_2\right)^2,\, D_7=-\left(p_2-l_1\right)^2,\, D_8=-\left(l_2-p_1\right)^2,\, D_9=-\left(l_2-p_2\right)^2\,,\nonumber
\end{gather}
where $d=4-2\epsilon$ is the space-time dimension, $\gamma_E=0.577\ldots$ is the Euler constant.
The first topology corresponds to the integrals with $n_{8,9}\leqslant 0$, while the second to those with $n_{7,8}\leqslant 0$. The IBP reduction reveals $47$ master integrals depicted in Fig. \ref{fig:masters}. 
We construct the differential systems with respect to $s=(p_1+p_2)^2$ and $t=(p_1-q_1)^2$: 
\begin{equation}
	\partial_s j=M_s(s,t,\epsilon) j\,,\quad \partial_t j=M_t(s,t,\epsilon) j\,,
\end{equation}
where $M_s$ and $M_t$ are the matrices with entries being rational in $s,\, t,$ and $\epsilon$.
 Then we apply the reduction algorithm from Ref. \cite{Lee2014} and reduce both differential systems to $\epsilon$-form using the private \textit{Mathematica} package \texttt{Libra}. In the process of reduction we pass to the variables $x$ and $z$ related to $s$ and $t$ via
\begin{equation}
	s/m^2=\frac{(x^2+1)^2}{x^2}\,,\quad t/m^2=-\frac{(x^2+1)^2}{x^2(z^2+1)}+1\,,
\end{equation}
which we discover partly heuristically and partly using the prescriptions of Ref. \cite{Lee2017a}.
These two variables are simply related to the velocity $\beta$ of the produced leptons in c.m.s. and the scattering angle $\theta$:
\begin{equation}
\beta =\frac{x^2-1}{x^2+1}\,,\quad \beta \cos\theta = \frac{z^2-1}{z^2+1}\,.
\end{equation}
In terms of variables $x,\,z$ the kinematic region corresponding to the annihilation channel is defined by the inequalities
\begin{equation}\label{eq:region}
	x>1\,,\quad x^{-1}\leqslant z \leqslant x\,.
\end{equation}
In what follows we will always assume that these inequalities hold. In particular, this allows us to treat $x$ as a plain notation for $\sqrt{\frac{1+\beta}{1-\beta}}$. 
The resulting differential systems can be written in $d\log$ form 
\begin{equation}\label{eq:dlog}
dJ = \epsilon dA J\,,
\end{equation}
where the canonical master integrals $J$ are related to $j$ via
\begin{equation}\label{eq:toespilon}
	j=TJ\,.
\end{equation}
Here $T=T(x,z)$ is a matrix with entries being the rational functions of $x$ and $z$, which can be found in the ancillary file \texttt{Transformation.m}. 
The matrix in the right-hand side of Eq. \eqref{eq:dlog} has the form
\begin{equation}
	dA = \sum_i A_i d\log P_i(x,z),
\end{equation}
$A_i$ being the numeric matrices, and the arguments $P_i$ of the logarithms belong to the following alphabet
\begin{multline}
	P_i\in \{x,\ x\pm1,\ x^2+1,\ x\pm z,\ x z\pm 1,\\
		x^4+x^2+1-x^2 z^2,\ x^2 z^2+x^4+3 x^2+1,\ (x^4+x^2+1) z^2-x^2\}.
\end{multline}
We fix the boundary conditions in two steps. First, we consider the backward direction determined by the curve
\begin{equation}
	xz=1.
\end{equation}
While this curve belongs to the singular locus of the differential equations, the specific solution that we are seeking is expected to be regular on it. Moreover, the boundary conditions on this curve are determined by the values of 31 corresponding integrals in backward kinematic ( \#\#1--19,22,24,26,27,30,31,34--36,39,43,45, see Fig. \ref{fig:masters}). We construct a dedicated differential system in $x$ (or rather in $\beta=\frac{x^2-1}{x^2+1}$) for the family of these integrals
\begin{equation}
	\partial_\beta\tilde j =\tilde M_\beta\tilde j\,,
\end{equation}
where $\tilde M_\beta$ is $31\times31$ matrix, depending rationally on $\beta$ and $\epsilon$, and  $$\tilde j=(j_1,\ldots,j_{19},j_{22},j_{24},j_{26},j_{27},j_{30},j_{31},j_{34},j_{35},j_{36},j_{39},j_{43},j_{45})^\intercal|_{\theta=\pi}$$ is a column vector of 31 master integrals in backward kinematic.
Then we reduce this differential system to $\epsilon$-form
\begin{gather}
	\partial_\beta\tilde J =\epsilon \tilde A(\beta)\tilde J\,,\\
	\tilde A(\beta)=\sum_{a\in\{0,\pm1,3\}}\frac{\tilde A_a}{\beta-a}\,,
	\label{eq:beta_eform}	
\end{gather}
where $\tilde J$ is a column of canonical master integrals related to $\tilde j$ via
\begin{equation}
	\tilde j = \tilde T \tilde J
\end{equation} 
with $\tilde T$ being rational in $\beta$ and $\epsilon$ matrix that we do not present here to save space.
It is remarkable that, in addition to singularities at $\beta=0,1,-1$, the above differential system has singularity at $\beta=3$.

The boundary conditions for this system are fixed from the asymptotics $m^2\to 0$ (or $\beta\to 1$). Namely, we calculate the following coefficients of this asymptotics:
%\begin{gather}
%	c_1(0),\,c_2(1-2 \epsilon ),\,c_3(0),c_4(1-2
%	\epsilon ),\,c_5(-\epsilon -1),\,c_6(1-\epsilon
%	),\,c_7(0),\,c_8(-\epsilon ),\,c_9(-2 \epsilon
%	),\,c_{10}(-\epsilon ),\nonumber\\
%	c_{11}(-2 \epsilon
%	),\,c_{12}(-2 \epsilon
%	),\,c_{13}(-1),\,c_{14}(-1),\,c_{15}(-\epsilon
%	),\,c_{16}(-2),\,c_{17}(\epsilon
%	),\,c_{18}(0),\,c_{19}(-1),\,c_{22}(-1),\nonumber\\
%	c_{24}(-4
%	\epsilon ),\,c_{26}(-1),\,c_{27}(-\epsilon
%	-2),\,c_{30}(2 \epsilon
%	-1),\,c_{31}(-1),\,c_{34}(-1),\,c_{35}(-1),\,c_{36}(-2-2\epsilon ),\nonumber\\
%	c_{39}(-1),\,c_{43}(-3 -2 \epsilon),\,c_{45}(-2),\label{eq:cs}
%\end{gather}
\begin{multline}
\tilde{c}^{\intercal}=(c_1^{2-2 \epsilon },c_2^{0},c_3^{1-2 \epsilon },c_4^{0},c_5^{1-\epsilon },c_6^{-\epsilon },c_7^{-2 \epsilon },c_8^{1-\epsilon },c_9^{0},c_{10}^{-\epsilon },c_{11}^{0},c_{12}^{0},c_{13}^{-2 \epsilon },c_{14}^{-2 \epsilon },c_{15}^{-\epsilon },c_{16}^{1-2 \epsilon },c_{17}^{-3 \epsilon },\\
c_{18}^{-2 \epsilon -1},c_{19}^{-2 \epsilon },c_{22}^{-2 \epsilon },c_{24}^{2 \epsilon -1},c_{26}^{-2 \epsilon },c_{27}^{-\epsilon },c_{30}^{-4 \epsilon },c_{31}^{-2 \epsilon -1},c_{34}^{-2 \epsilon },c_{35}^{-2 \epsilon -1},c_{36}^{0},
c_{39}^{-2 \epsilon -1},c_{43}^{0},c_{45}^{-2 \epsilon -1}\big)\,,\\ \label{eq:cs}
\end{multline}
where $c_k^\alpha$ denotes the coefficient in front of $(1-\beta)^\alpha$ in the small-mass asymptotics of the $k$-th integral. We succeed to calculate all constants, but the last three, exactly in $\epsilon$ in terms of the hypergeometric functions ${}_{q+1}F_q$ which can be expanded using the \texttt{HypExp} package \cite{Huber2008a} virtually to any order in $\epsilon$. For this purpose we use the \texttt{asy} program \cite{Pak:2010pt} and determine the relevant regions in parametric representation. Note that the small-mass asymptotics contains explicit logarithms already before the expansion in $\epsilon$. This signals the necessity to introduce, in addition to dimensional, the analytical regularization and we do it in many cases. The analytical regularization can then be removed without introducing the derivatives of hypergeometric functions with respect to indices, with one exception of $c_{34}^{-2\epsilon}$. For the latter constant the removal of the analytic regularization required  derivatives of hypergeometric functions. However, after the expansion in $\epsilon$ these derivatives are taken at integer values of indices of the hypergeometric functions and, therefore, expressed in terms of the conventional constants. Note that the constants with zero argument correspond to zero-mass limit of the integrals. In particular, $c_{36}^0$ is an on-shell massless vertex integral known from Ref. \cite{Gehrmann:2005pd}.

The last three constants $c_{39}^{-2 \epsilon -1},c_{43}^0$, and $c_{45}^{-2 \epsilon -1}$ are fixed by applying constraints coming from different singular points. Namely, the constant $c_{39}^{-2 \epsilon -1}$ is obtained from the condition of absence of the term $\propto \beta^{2\epsilon}$ in the $\beta\to 0$ asymptotics of the integral \#39, while the two constants $c_{43}^0$ and $c_{45}^{-2 \epsilon -1}$ are fixed from the condition of absence of the terms $\propto (1+\beta)^{-2}$ and  $\propto (1+\beta)^{-1+2\epsilon}$ in the $\beta\to -1$ asymptotics of integrals \#43 and \#45, respectively. Since the evolution operators connecting the point $\beta=1$ and the points $\beta=0,-1$ is known only as the expansion in $\epsilon$, our results for these three constants also have the form of $\epsilon$-expansion. Altogether, the expanded in $\epsilon$ results for all constants from Eq. \eqref{eq:cs} are expressed in terms of multiple zeta values with positive and negative indices, at least, up to transcendentality weight 6.

\section{Results}

\begin{figure}
	\centering
	\includegraphics[width=0.5\linewidth]{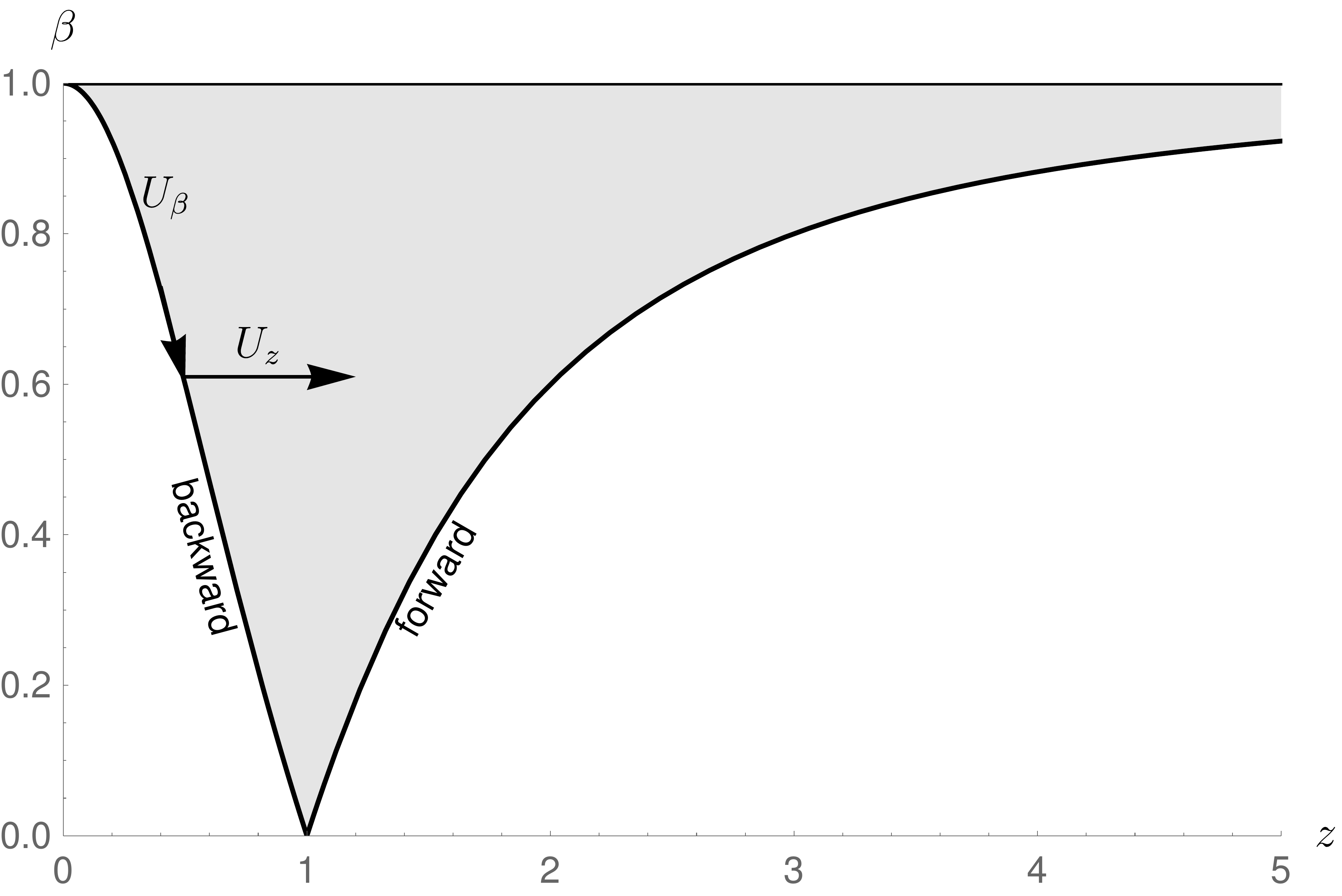}
	\caption{Physical region in variables $\beta$ and $z$. The operator $U_\beta$ determines the evolution along the left boundary of the physical region, while $U_z$ determines the evolution in $z$ direction. \label{fig:region}}
\end{figure}

We write the specific solution of the differential system \eqref{eq:dlog} in the form
\begin{equation}
	J= U_z L U_{\beta} \tilde{L} \tilde c\,,
\end{equation}
where $\tilde c$ is defined in Eq. \eqref{eq:cs}, $U_{\beta}$ and $U_{z}$ are the evolution operators (see Fig. \ref{fig:region})
\begin{align}
	U_{\beta}&=\Pexp[\epsilon\intop_1^{\beta} d\beta^\prime\tilde A(\beta^\prime)]\nonumber\,,\\
	U_{z}&=\Pexp[\epsilon\intop_{1/x}^{z} dz^\prime \frac{dA}{dz^\prime}(x,z^\prime)]\,,
\end{align}
$\tilde L$ is a $31\times 31$ matrix with entries being the rational functions of $\epsilon$, and $L$ is a $47\times 31$ purely numerical matrix. The meaning of the matrix $\tilde L$ is that it determines the relation between the coefficients $\tilde c$ in the small-mass asymptotics of the backward integrals and the boundary constants $\widetilde C$ for the solution $\tilde J=U_{\beta}\widetilde C$ of Eq. \eqref{eq:beta_eform}. The meaning of the matrix $L$ is very similar: it determines the relation between the integrals $\tilde J$ and the boundary constants $C$ in the solution  $J=U_{z}C$ of Eq. \eqref{eq:dlog}. Note that in what follows we always put final particles mass to one, $m=1$.

The evolution operators $U_\beta$ and $U_z$ can be evaluated in $\epsilon$ expansion and involve Goncharov's polylogarithms $G$, Ref. \cite{goncharov1995polylogarithms}. The operator  $U_\beta$ involves polylogarithms of the form $G(a_n,\ldots, a_1|1-\beta)$, where the letters $a_k$ belong to the alphabet $\{0,1,2,-2\}$. By a direct inspection one can check that the letters $1$ and $-2$ never appear simultaneously. This allows us to express all polylogarithms in $U_\beta$ via harmonic polylogarithms. We do it as follows. For the polylogarithms without letter $-2$ we pass from the argument $1-\beta$ to the argument $\beta$. For the polylogarithms without letter $1$ we pass from the argument $1-\beta$ to the argument $(1-\beta)/2$. After these transformations the expansion of $U_\beta$ is expressed via harmonic polylogarithms of the arguments $\beta$ and $(1-\beta)/2$.

The complexity of the results mostly comes from the operator $U_z$. It is convenient to temporarily express its $\epsilon$ expansion in terms of the iterated integrals 
\begin{equation}\label{eq:II}
	\II(l_n,\ldots l_1|z)=\intop_{1/x<z_1<\ldots <z_n<z}\hspace{-1cm} dz_1\ldots dz_n\,l_1(z_1)\ldots l_n(z_n)\,,
\end{equation}
where each $l_k(z)$ is one of the nine weights
\begin{gather}
w_1(z) = \frac{1}{z},\ w_2(z)=\frac{2 z}{z^2-x^2},\ w_3(z) = \frac{2 z}{z^2-x^{-2}},\ w_4(z) = \frac{2 z}{z^2+1},\nonumber\\
w_5(z) = \frac{2 z}{z^2 -(x^2+1+x^{-2})},\ 
w_6(z) = \frac{2 z}{z^2+x^2+3+x^{-2}},\ w_7(z) = \frac{2 z}{z^2-(x^2+1+x^{-2})^{-1}},\nonumber\\
m_2(z) = \frac{2 x}{z^2-x^2},\ m_3(z) = \frac{2 x^{-1}}{z^2-x^{-2}}\,.
\end{gather}
These weights can be partial-fractioned into standard weights of the form $1/(z-z_k)$ where $z_k$ is one of
\begin{equation*}
	z_0=0,\,z_{1-4}=\pm x^{\pm 1},\,z_{5,6}=\pm i,\,z_{7-10}=\pm(x^2+1+x^{-2})^{\pm 1/2},\,z_{11,12}=\pm i (x^2+3+x^{-2})^{1/2} \,.
\end{equation*}
Therefore, the $\epsilon$ expansion of $U_z$ can be expressed via $G(\ldots |z-x^{-1})$ with letters of the form $a_k=z_k-x^{-1}$ ($k=0,1,\ldots,12$). The shift of the argument and the letters by $x^{-1}$ is due to the lower integration limit $x^{-1}$ in the appearing iterated integrals \eqref{eq:II}. 

However we find that in all sectors except two non-planar ones the weights $m_{2,3}$ do not appear. This allows us to express the results in these sectors via $G(\ldots|z^2-x^{-2})$ with letters of the form $b_k=u_k-x^{-2}$, where $u_k$ is one of
\begin{equation*}
u_1=0,\,u_{2,3}=x^{\pm 1},\,u_4=1,\,u_{5,6}=(x^2+1+x^{-2})^{\pm 1},\,u_7=-(x^2+3+x^{-2})\,.
\end{equation*}
The exception are the two non-planar sectors containing 7 master integrals \#\#39 --42,45--47. The results in these sectors contain iterated integrals depending on the weights $m_{2,3}$ and we  express them via $G(\ldots |z-x^{-1})$. Remarkably, we find that the letters $a_{11,12}$ never appear in our results. We would like to stress that both $U_\beta$ and $U_z$ are real quantities in the whole physical region \eqref{eq:region} by construction since the weights $w_{1-7},\,m_{2,3}$ are real and finite. Therefore, the imaginary parts of the master integrals appear only due to the imaginary parts of boundary constants $\tilde{c}$.

In the ancillary file \texttt{Jresults.m} we present the results for the canonical master integrals $J$. The original master integrals $j$ are related to the canonical ones via $j=TJ$, where $T$ is the transformation matrix as defined in the attached file \texttt{Transformation.m}. We find this way of presenting the results more preferable then just presenting one file with results for the original master integrals $j$. The reason is that, when multiplying $T$ and $J$,  we can loose some terms of high transcendental weights (in particular, the 4th t.w.) due to the truncation of power series in $\epsilon$. Meanwhile, having results for $T$ and $J$ separately allows us to first represent the amplitudes in terms of the canonical master integrals without any expansion in $\epsilon$ and to substitute the expansion of the canonical master integrals only on the last step. This way we secure that all terms of up to the highest transcendental weight kept in the expansion of $J$ are retained in the amplitude.

\section{$x\to x^{-1}$ symmetry and limiting cases.}

The set of 31 integrals in backward direction appears to be very convenient for analyzing both threshold and forward limits. While for the threshold limit this is clear from Fig. {\ref{fig:region} as we can travel along the left boundary to the point of intersection with the $z$ axis, the forward limit deserves some discussion. The integrals in forward kinematic can be obtained by the formal change of sign of $\beta$. However, in general, this change of sign is not that trivial since, in order to pass to the region of negative $\beta$, one has to bypass the singular point $\beta=0$. Fortunately, our set of 31 integrals is analytic at $\beta=0$. This can be checked by examining the possible fractional powers of $\beta$ in the general solution of the differential system \eqref{eq:beta_eform}. They all appear to be of the form $k\epsilon$ with positive integer $k$. These powers can not appear in the expansion of the integrals as can be understood using expansion by regions\footnote{The physical reason for the absence of the non-analytic terms at the threshold is that the diagrams that we consider do not allow for soft photon exchange between two final particles. We thank Andrei Grozin for bringing this to consideration.}. Therefore, the two contours, bypassing the point $\beta=0$ from above and from below, are equivalent. This fact is explicit in our results as they  do not contain harmonic polylogarithms of the argument $\beta$ with trailing zeros.

Let us write this symmetry as
\begin{equation}
	j(x,z=x) = j(x,z=x^{-1})|_{x\to x^{-1}}
\end{equation}
The left-hand side of this equation are the integrals in forward direction, while the right-hand side are the integrals in backward direction in which the replacement $\beta\to -\beta$ is made. The right-hand side can be rewritten as $ j(x^{-1},z=x)$. In fact, the above symmetry can be promoted to the one for arbitrary $z$:
\begin{equation} \label{eq:xsymmetry}
j(x,z) = j(x^{-1},z)
\end{equation}
In terms of the variables $\beta$ and $\cos\theta$ this symmetry corresponds to the change
\begin{equation}
\beta\to -\beta,\quad \cos\theta\to -\cos\theta\,.
\end{equation}
Again, we stress that this naive change of the sign of $\beta$ is justified only due to the analyticity of the integrals on the threshold $\beta=0$ and would be illegal for the diagrams with the photon exchange between final particles.

Let us now obtain the relation for the canonical master integrals. Substituting $j=T J$ in Eq. \eqref{eq:xsymmetry} we obtain
\begin{equation}
J(x,z) = D J(x^{-1},z)\,,
\end{equation}
where $D=[T(x,z)]^{-1} T(x^{-1},z)$. Remarkably, we find that $D$ is purely numerical matrix independent of $x$ and $z$. This matrix is diagonalizable, with eigenvalues being mostly $+1$ except the five eigenvalues equal to $-1$. Having learned that, we have adjusted the transformation $T$ from Eq. \eqref{eq:toespilon} so that $D$ is explicitly diagonal, with $-1$ standing on the main diagonal in the positions $10,\, 31,\, 34,\, 37,\, 40$. Note that while doing this adjustment we have kept the consistency of our transformation $T$ with the sectorial hierarchy of the master integrals.
Finally, we obtain 
\begin{equation}\label{eq:symmetry}
J_k(x,z) = \sigma_k J_k(x^{-1},z)\,,
\end{equation}
where 
\begin{equation}
	\sigma_k=\begin{cases}
		-1\quad k\in\{10,31,34,37,40\}\,,\\ 
		+1\quad \text{otherwise}\,.
		\end{cases}
\end{equation}
Note that our results are not explicitly (anti)symmetric in $x\to x^{-1}$. Therefore, we have used this symmetry to perform the rigorous consistency check of our results by numerically evaluating both sides of the identity \eqref{eq:symmetry} using \texttt{GiNaC}\footnote{After the replacement $x\to x^{-1}$ the arguments of the Goncharov's polylogarithms become $z-x$ and $z^2-x^2$, so they are both negative in the physical region \eqref{eq:region}. Therefore, in order to use \texttt{GiNaC} we used the property $G(a_n,\ldots ,a_1|x)=G(-a_n,\ldots, -a_1|-x)$ which holds for $a_1\neq 0$.}.

\paragraph*{Backward/Forward limits.}
The expressions for the integrals presented in the ancillary file \texttt{Jresults.m} depend on $z$ via the Goncharov's polylogs of the form $G(\ldots |z-x^{-1})$ and $G(\ldots |z^2-x^{-2})$ with no trailing zeros. Therefore, in order to obtain the backward limit we simply have to replace all $G$ by zero:
\begin{equation}\label{eq:backward}
J_k(\cos\theta = -1) = J_k(x,z=x^{-1}) = J_k(x,z)|_{G\to 0}\,.
\end{equation}

Forward limit is easily obtained from the integrals in backward limit. One simply has to use Eq. \eqref{eq:symmetry} to obtain
\begin{equation}\label{eq:forward}
J_k(\cos\theta = 1) = \sigma_k J_k(\cos\theta = -1)|_{\beta\to -\beta}
\end{equation}
We do not present separately the results for forward and backward limits as the transformations \eqref{eq:backward} and \eqref{eq:forward} can be trivially applied to the content of the file \texttt{Jresults.m}. Note that the transformation $T$ from the file \texttt{Transformation.m} is regular at $z=x^{\pm1}$ and therefore the forward/backward limit of the original master integrals is also trivially recovered.

\paragraph*{Threshold limit.} Although the canonical master integrals have finite threshold limit, the transformation matrix $T(x,z)$ is singular at $x=1$. Therefore we prefer to present the results for the threshold values of the initial master integrals $j$. In these results we choose to keep the terms up to transcendentality weight $5$, which should be sufficient for the applications.
The results are presented in the ancillary file \texttt{jThresholdResults.m}. We have managed to express all harmonic polylogarithmic constants with transcendentality weight $\leq 4$ in terms of constants which can be evaluated by \textit{Mathematica} without any additional packages. For the transcendentality weight $5$ we have in addition the three polylogarithmic constants $H(2,-3|\frac12)$, $H(3,-2|\frac12)$, and $H(-3,-2|\frac12)$. To give the reader an impression of how the threshold results look like, we present here two most complicated ones truncated to the terms of transcendentality weight $4$.
\begin{multline}
(2 \epsilon -1) (3 \epsilon -1) j_{12}=\frac{1}{2 \epsilon ^2}-\frac{2 \ln 2}{\epsilon }+\frac{\text{Li}_3\left(-\tfrac{1}{2}\right)}{2}+\frac{1}{2} \text{Li}_2\left(-\tfrac{1}{2}\right) \ln 2+\frac{21 \zeta (3)}{16}+\frac{\ln ^32}{6}+3 \ln ^22\\
-\frac{1}{4} \ln ^22 \ln 3-\frac{1}{4} \pi ^2 \ln 3+\epsilon  \bigg(-\frac{5}{6} \pi ^2 \text{Li}_2\left(-\tfrac{1}{2}\right)-3 \text{Li}_3\left(-\tfrac{1}{2}\right)+\frac{3}{2}\text{Li}_4\left(-\tfrac{1}{3}\right)+\frac{14 }{3}\text{Li}_4\left(\tfrac{1}{2}\right)+6 \text{Li}_4\left(\tfrac{2}{3}\right)\\
+\frac{3}{2}\text{Li}_4\left(\tfrac{3}{4}\right)
-2 \text{Li}_2\left(-\tfrac{1}{2}\right) \ln ^22-3 \text{Li}_2\left(-\tfrac{1}{2}\right) \ln 2-5 \text{Li}_3\left(-\tfrac{1}{2}\right) \ln 2-\frac{251 \zeta (3)}{24}-\frac{7}{24} \zeta (3) \ln 2-\frac{619 \pi ^4}{4320}\\
+\frac{37 \ln ^42}{36}+\frac{5 \ln ^43}{16}-\frac{13 \ln ^32}{3}-\frac{3}{2} \ln ^32 \ln 3-\ln 2 \ln ^33-\frac{10}{9} \pi ^2 \ln ^22+\frac{3}{2} \ln ^22 \ln 3-\frac{3}{8} \pi ^2 \ln ^23\\
+\frac{3}{2} \ln ^22 \ln ^2(3)-\frac{5}{6} \pi ^2 \ln 2+\frac{3}{2} \pi ^2 \ln 3+\frac{4}{3} \pi ^2 \ln 2 \ln 3\bigg)+O\left(\epsilon ^2\right),
\end{multline}
\begin{multline}
j_{45}=-\frac{5}{96 \epsilon ^4}+\frac{-\frac{7}{96}-\frac{3 i \pi }{64}+\frac{5 \ln 2}{24}}{\epsilon ^3}+\frac{\frac{7}{24}-\frac{i \pi }{8}+\frac{11 \pi ^2}{192}-\frac{5 \ln ^22}{12}+\frac{7 \ln 2}{24}+\frac{3}{16} i \pi  \ln 2}{\epsilon ^2}\\
+\frac{\frac{211 \zeta (3)}{288}-\frac{7}{6}+\frac{i \pi }{2}-\frac{\pi ^2}{192}+\frac{9 i \pi ^3}{128}+\frac{5 \ln ^32}{9}-\frac{7 \ln ^22}{12}-\frac{3}{8} i \pi  \ln ^22-\frac{7 \ln 2}{6}+\frac{1}{2} i \pi  \ln 2-\frac{11}{48} \pi ^2 \ln 2}{\epsilon }\\
-3 \text{Li}_4\left(\tfrac{1}{2}\right)-\frac{395 \zeta (3)}{144}+\frac{61 i \pi  \zeta (3)}{32}-\frac{337}{72} \zeta (3) \ln 2+\frac{14}{3}-2 i \pi +\frac{\pi ^2}{48}+\frac{i \pi ^3}{16}+\frac{23 \pi ^4}{1280}-\frac{49 \ln ^42}{72}+\frac{7 \ln ^32}{9}\\
+\frac{1}{2} i \pi  \ln ^32+\frac{7 \ln ^22}{3}-i \pi  \ln ^22+\frac{7}{12} \pi ^2 \ln ^22+\frac{14 \ln 2}{3}-2 i \pi  \ln 2-\frac{29}{48} \pi ^2 \ln 2-\frac{7}{32} i \pi ^3 \ln 2+O\left(\epsilon ^1\right)
\end{multline}

%\section{Checks}

\section{Discussion and Conclusion}

In the present paper we have calculated the family of the master integrals relevant for the NNLO correction to the process $e^+e^-\to\ell^+\ell^-$ with the full account of the mass of the final particles. This family is related to the two big topologies with three-photon intermediate state. We stress here that our results are applicable directly to the physical region corresponding to the annihilation channel with imaginary parts originating only from the boundary constants. Our main results are presented in the ancillary files \texttt{Jresults.m} and \texttt{Transformation.m}, the first containing the results for the canonical master integrals and the second the transformation matrix to the original master integrals. For the reader convenience, we attach also the Mathematica notebook file \texttt{Numerics.nb} which contains the code for numerical evaluation of the master integrals.

We have performed various checks of our results. Most of the integrals have been checked using \texttt{Fiesta} \cite{Smirnov:2015mct}. However, for the integrals \#\#45-47 in the most complicated sector this check appears to be hardly feasible. Here we have followed the same lines as in Ref. \cite{DiVita2018}. Namely, we have used the fact that the integral \#45 is finite at $d=6$ and can be calculated numerically, even including the higher in $\epsilon=3-d/2$ terms. Then, using the dimensional recurrence relations we have represented $j_{45}|_{d=6-2\epsilon}$ as a linear combination of $j_{k}|_{d=4-2\epsilon}$ and checked the available expansion terms to find the perfect agreement. Note that this check is very rigorous as almost all master integrals contributed to this linear combination (except the three integrals \#\#11,43,44 which are related solely to planar topology). In contrast to the approach of Ref. \cite{DiVita2018}, we did not do any analytical integrations but rather relied on the \textit{Mathematica} \texttt{NIntegrate} routine. However, since the second Symanzik polynomial $F$ vanishes inside the integration region we had to modify the integration contours. Let us explain some details. We start from the Feynman parametrization for the integral $j_{45}$
\begin{equation}\label{eq:FP}
j_{45}(x,z)|_{d=6-2\epsilon} = e^{2\epsilon \gamma_E}\Gamma (7-d)\intop_{0}^{\infty}\ldots  \intop_{0}^{\infty}\frac{dx_1\ldots dx_7}{U^{3d/2-7}F^{7-d}}\delta(1-\sum x)\,,
\end{equation}
\begin{align*}
U=&x_1 x_{2457}+ x_2x_{3567}+x_4 x_{36}+x_{57}x_{346},\\
F=&2 x_1 x_2 x_{57}+x_1^2x_{2457}+x_2^2 x_{13567}+(1-u)x_2 x_3 x_7\\
&+(s-1+u) \left(x_{2457}x_1x_6+x_1x_4x_7+x_2 x_5 x_6\right)-s x_3 x_4 x_5-i0\,,
\end{align*}
where $u=2-s-t$, $x_{ij\ldots k}=x_i+x_j+\ldots x_k$ and the sum in the argument of the $\delta$-function goes over any nonempty subset of $\{1,\ldots,7\}$. We assume that $d=6-2\epsilon$ and interpret the sum in the argument of the $\delta$-function as $x_{12}$. Note that this choice leads to the integration from $0$ to $\infty$ over $x_{3,\ldots,7}$. Since $s+t-1=1-u>0$ and $1-t>0$, the only negative term in $F$ polynomial is $-s x_3 x_4 x_5$. Therefore, in order to avoid zeros of $F$, we can slightly rotate the contours of integration over $x_6$ and $x_7$ in a clockwise direction. We do it by replacing 
\begin{equation}
x_{6,7}\to e^{-i \pi/3} x_{6,7}
\end{equation} 
This simple transformation makes the integral absolutely convergent. Then the expansion in $\epsilon$ can be performed under the integral sign. We then use the \textit{Mathematica} routine \texttt{NIntegrate} to perform numerically the 6-fold integral. In less then 2 minutes we find
\begin{equation}
	j_{45}(x=2,z=\tfrac23)|_{d=6-2\epsilon} = 0.3829 +0.2319 i +(0.445 + 1.162 i)\epsilon +(1.0 + 3.05 i)\epsilon^2 +O(\epsilon)^3.
\end{equation}
This is to be compared with 
\begin{equation}
	j_{45}(x=2,z=\tfrac23)|_{d=6-2\epsilon} = (0.38309\ldots
	+0.23187\ldots i)+(0.4457\ldots+1.1624\ldots i) \epsilon +O\left(\epsilon ^2\right),
\end{equation}
that we have obtained from our results using dimensional recurrence relations.

There are also several self-consistency checks that we have made. The first, and the most rigorous check is the one related to $x\to x^{-1}$ symmetry. Then there is a check related to the threshold integrals. We have calculated the threshold values of $47$ master integrals in Fig. \ref{fig:masters}. However, there are only $22$ threshold master integrals. Therefore we have had $47-22=25$ consistency checks.

Finally, we have selectively compared some of our results with those of Ref. \cite{DiVita2018}. As the question of the analytic continuation is not trivial, we tried to empirically guess the proper prescription. The best guess was the replacement rule
\begin{equation}\label{eq:subst1}
	w=w_{\text{Ref.\cite{DiVita2018}}} \to  -x^2-i0\,, \quad  \tilde z = z_{\text{Ref.\cite{DiVita2018}}} \to  z x\,.
\end{equation}
which seems to reproduce real parts, and also the imaginary parts, up to the opposite sign. One might think of a trivial typo, however the situation turns out to be more interesting. Let us consider the 
integral denoted as $\mathrm{I}_{42}$ in Ref. \cite{DiVita2018}. The two initial terms of $\epsilon$ expansion of $\mathrm{I}_{42}$ from Ref. \cite{DiVita2018} have the form:
\begin{equation}\label{eq:LM1}
\mathrm{I}_{42,\, \text{Ref.\cite{DiVita2018}}} = -\frac34 + \epsilon\left(-\frac{3}{2} \log \left(1-\frac{w}{\tilde{z}^2}\right)-\frac{\log \left(\tilde{z}\right)}{3}+\frac{9}{2} \log (1-w)-\frac{25 \log (w)}{12}-\frac{4 i \pi}{3}\right) + O(\epsilon)^2 
\end{equation}
while our result reads
\begin{align}\label{eq:L}
\mathrm{I}_{42} &=
 -\frac34 + \epsilon\left(\frac{9}{2} \log \left(x^2+1\right)-\frac{9 \log (x)}{2}-\frac{3}{2} \log \left(z^2+1\right)+\frac{8 \log (z)}{3}-\frac{3 i \pi}{4}\right) + O(\epsilon)^2 \nonumber \\
 &=-\frac34 + \epsilon\left(\frac{4}{3} \log (1-u)+\frac{3 \log (s)}{4}+\frac{1}{6} \log (1-t)-\frac{3 i \pi}{4}\right) + O(\epsilon)^2
\end{align}
Indeed, we see that the substitution \eqref{eq:subst1} in Eq. \eqref{eq:LM1} leads to the imaginary part ${25i\pi}/{12} - {4i\pi}/{3}=+{3i\pi}/4$ which has the sign opposite to the correct one. Let us however express \eqref{eq:LM1} via $s$ and $t$ using 
\begin{equation}
s_{\text{Ref.\cite{DiVita2018}}}= t,\quad 
t_{\text{Ref.\cite{DiVita2018}}}= s\,.
\end{equation}
We have
\begin{equation}\label{eq:LM2}
\mathrm{I}_{42,\, \text{Ref.\cite{DiVita2018}}} = -\frac34 + \epsilon\left(\frac{4}{3} \log (u-1)+\frac{3 \log (-s)}{4}+\frac{1}{6} \log (1-t)-\frac{4 i \pi}{3}\right) + O(\epsilon)^2
\end{equation}
This result is unambiguous in its dedicated region, $s<0\&t<0$. Now we see that the correct analytical continuation to the annihilation channel is obtained if we rewrite it as 
\begin{equation}\label{eq:LM3}
\mathrm{I}_{42,\, \text{Ref.\cite{DiVita2018}}} = -\frac34 + \epsilon\left(\frac{4}{3} \log (1-u-i0)+\frac{3 \log (-s-i0)}{4}+\frac{1}{6} \log (1-t-i0)\right) + O(\epsilon)^2
\end{equation}
This corresponds to a natural prescription
\begin{equation}\label{eq:subs2}
	s\to s+i0\,,\quad t\to t+i0\,,\quad u\to u+i0\,.
\end{equation}
Then in the annihilation channel we have $\log (1-u-i0)=\log (1-u)\,,\ \log (-s-i0)=\log(s)-i\pi\,,\ \log (1-t-i0)=\log (1-t)$ so that the imaginary part becomes $-3i\pi/4$ which agrees with Eq. \eqref{eq:L} including the sign. 

While we could guess the proper prescription for the logarithms, it is quite nontrivial to do the same for the more complex polylogarithms. The problem is that $s,t,$ and $u$ are not independent variables and therefore our prescription \eqref{eq:subs2} lacks the exact mathematical meaning. This kind of problems is connected with the fact that it is not possible to find the analogue of the Euclidean region on the plane $s,t$, i.e., the region where the polynomial $F$ in the Feynman parametrization \eqref{eq:FP} is positive-definite for the whole integration region over Feynman parameters. One way out of this would be putting one of the final particles off-shell so that $u$ becomes independent variable. However this would clearly be an overshoot in terms of efforts.

\acknowledgments
R.L. is grateful to Andrei Grozin and Andrei Pomeransky for useful discussions and to Vladimir Smirnov for the help with \texttt{Fiesta} checks. This work is supported by RFBR grant 17-02-00830.

\bibliographystyle{JHEP}
%\bibliography{include/eetautau}

\providecommand{\href}[2]{#2}\begingroup\raggedright\endgroup

\end{document}